\documentstyle[12pt,aaspp4,flushrt]{article}
\def\fig #1, #2, #3 {
  \smallskip
  \centerline{\psfig{figure=#1,height=#2 in,width=#3 in}} }
\def\capt{\small \baselineskip 12pt }

\def\equ#1{equation~(\ref{eq:#1})}
\def\eqd#1{eq.~[\ref{eq:#1}]}
\def\se#1{\S\ref{sec:#1}}
\def\Figu#1{Figure~\ref{fig:#1}}

\def\cl{\centerline}
\def\\{\hfill\break}

\def\etal{{\it et al.\ }}

\def\eg{{e.g.}}
\def\ie{{i.e.}}

\def\vev#1{\langle#1\rangle}
\def\av{\vev}

\def\be{\begin{equation}}
\def\ee{\end{equation}}
\newcommand{\brr}{\begin{array}}
\newcommand{\err}{\end{array}}

\def\ifm#1{\relax\ifmmode#1\else$\mathsurround=0pt #1$\fi}
\def\kms{\ifmmode\,{\rm km}\,{\rm s}^{-1}\else km$\,$s$^{-1}$\fi} 
\def\hmpc{\,\ifm{h^{-1}}{\rm Mpc}}
\def\dd{d}

\def\msolar{M_{\odot}}
\def\hmsun{h^{-1}\msolar}

\def\ltsima{$\; \buildrel < \over \sim \;$}
\def\lsim{\lower.5ex\hbox{\ltsima}}
\def\gtsima{$\; \buildrel > \over \sim \;$}
\def\gsim{\lower.5ex\hbox{\gtsima}}
 
\def\pmb#1{\setbox0=\hbox{#1}%
 \kern-.025em\copy0\kern-\wd0
 \kern.05em\copy0\kern-\wd0
 \kern-.025em\raise.0433em\box0}

\def\vx{\pmb{$x$}}

\def\om{\Omega_{\rm m}}
\def\ol{\Omega_\Lambda}
\def\bh{\hat{b}}
\def\bt{\tilde{b}}

\def\bha{\hat{b}_{\rm a}}
\def\bta{\tilde{b}_{\rm a}}
\def\sigb{\sigma_{\rm b}}
\def\delg{\delta_{\rm g}}
\def\delgz{\delta_{\rm g,z}}
\def\delmz{\delta_{\rm z}}

\def\dellnz{\delta_{\rm ln,z}}
\def\delgone{\delta_{\rm g_1}}
\def\delgtwo{\delta_{\rm g_2}}
\def\delm{\delta}
\def\del{\delta}
\def\delh{\delta_{\rm g}}
\def\coav{\langle\delg \vert\delm\rangle}
\def\coavh{\langle\delg \vert\delm\rangle}
\def\sigm{\sigma}
\def\sig{\sigma}
\def\sigmz{\sigma_{\rm z}}
\def\sigg{\sigma_{\rm g}}
\def\sigh{\sigma_{\rm g}}
\def\cg{C_{\rm g}}
\def\cgz{C_{\rm g,z}}

\def\cgone{C_{\rm g_1}}
\def\cgtwo{C_{\rm g_2}}
\def\cm{C}
\def\cmz{C_{\rm z}}
\def\ch{C_{\rm g}}
\def\pln{P_{\rm ln}}
\def\cln{C_{\rm ln}}
\def\clnz{C_{\rm ln,z}}

\def\neff{N_{\rm eff}}
\def\volbox{V_{\rm box}}
\def\vwin{V_{\rm win}}

\begin{document}
\medskip
\baselineskip 14pt 

\title{Measuring the Nonlinear Biasing Function\\
       from a Galaxy Redshift Survey}
 
\author{Yair Sigad\altaffilmark{1}, 
Enzo Branchini\altaffilmark{2},
\& Avishai Dekel\altaffilmark{1} }

\altaffiltext{1}{Racah Institute of Physics, The  Hebrew University,
Jerusalem 91904, Israel}
\altaffiltext{2}{Kapteyn Institute, University of Groningen,
Landleven 12, 9700 AV, Groningen, The Netherlands}

\begin{abstract}

We present a simple method for evaluating the nonlinear biasing function of 
galaxies from a redshift survey.  The nonlinear biasing is characterized by 
the conditional mean of the galaxy density fluctuation given the underlying 
mass density fluctuation $\coav$, or by the associated parameters of mean 
biasing $\bh$ and nonlinearity $\bt$ (following Dekel \& Lahav 1999).  Using 
the distribution of galaxies in cosmological simulations, at smoothing of a 
few Mpc, we find that $\coav$ can be recovered to a good accuracy from the 
cumulative distribution functions of galaxies and mass, $\cg(\delg)$ and 
$\cm(\delm)$, despite the biasing scatter.
Then, using a suite of simulations of different cosmological models, we 
demonstrate that $\cm(\delm)$ can be approximated in the mildly nonlinear 
regime by a cumulative log-normal distribution of $1+\delm$ with a single 
parameter $\sigm$, with deviations that are small compared to the difference 
between $\cg$ and $\cm$.
Finally, we show how the nonlinear biasing function can be 
obtained with adequate accuracy directly from the observed $\cg$ in redshift 
space. Thus, the biasing function can be obtained from counts in cells once the
rms mass fluctuation at the appropriate scale is assumed a priori. The relative
biasing function between different galaxy types is measurable in a similar way.
The main source of error is sparse sampling, which requires that the mean 
galaxy separation be smaller than the smoothing scale. Once applied to redshift
surveys such as PSC$z$, 2dF, SDSS, or DEEP, the biasing function can
provide valuable  
constraints on galaxy formation and structure evolution.

\end{abstract}

\subjectheadings{cosmology: theory --- cosmology: observation ---
dark matter --- galaxies: distances and redshifts --- 
galaxies: formation --- galaxies: clustering ---
large-scale structure of universe}

\vfill\eject

\section{INTRODUCTION}
\label{sec:intro}

The fact that galaxies of different types cluster differently 
(\eg, Dressler 1980; Lahav, Nemiroff \& Piran 1990; Santiago \& Strauss 1992; 
Loveday \etal 1995; Hermit \etal 1996; Guzzo \etal 1997) 
indicates that the galaxy distribution is in general biased compared 
to the underlying mass distribution.  
Cosmological simulations confirm that halos and galaxies 
must be biased (\eg, Cen \& Ostriker 1992; Kauffmann, Nusser \&
Steinmetz 1997; Blanton \etal 1999; Somerville \etal 2000).
The biasing becomes even more pronounced at high redshift,
as predicted by theory (\eg, Kaiser 1986; Davis \etal 1985;
Bardeen \etal 1986; Dekel \& Rees 1987; Mo \& White 1996; 
Bagla 1998; Jing \& Suto 1998; Wechsler \etal 1998), 
and confirmed by the strong clustering of galaxies observed
at $z\sim 3$ (Steidel \etal 1996; 1998).
Knowing the biasing scheme is crucial for extracting dynamical
information and cosmological constants from the observed galaxy distribution,
and may also be very useful for understanding the process and history
of galaxy formation.
 
The simplest possible biasing model relating the density fluctuation
fields of matter and galaxies, $\delm$ and $\delg$,
is the deterministic and linear
relation, $\delg(\vx)=b\,\delta(\vx)$, where $b$ is a constant linear
biasing parameter. However, this is at best a crude approximation,
because it is not self-consistent (\eg, it does not prevent $\delg$ from
becoming smaller than $-1$ when $b>1$) and is not preserved in time.
At any given time, scale and galaxy type, the biasing is expected in general 
to be nonlinear, i.e., $b$ should vary as a function of $\delta$.
The nonlinearity of dark-matter halo biasing (as well as its dependence 
on scale, mass and time) is approximated fairly well by the model of 
Mo \& White (1996), based on the extended Press-Schechter formalism 
(Bond \etal 1991).
Improved approximations have been proposed by 
Jing (1998), Catelan \etal (1998), Sheth \& Tormen (1999) and Porciani
\etal (1999). 
It is quantified further for halos and galaxies  
using cosmological $N$-body simulations with semi-analytic
galaxy formation (\eg, Somerville \etal 2000).
The biasing is also expected, in general, to be stochastic, in the sense that
a range of values of $\delg$ is possible for any given value of $\delm$.
For example, if the biasing is nonlinear on one scale, 
it should be different 
and non-deterministic on any other scale.
The origin of the scatter is shot noise as well as the influence
of physical quantities other than mass density (\eg, velocity
dispersion, 
the dimensionality of the local deformation tensor which affects 
the shape of the collapsing object,
etc.) on the efficiency
of galaxy formation.

Dekel \& Lahav (1999) have proposed a general formalism
for galaxy biasing, that separates nonlinearity and stochasticity in a
natural way. The density fields are treated as random fields, and the
biasing is fully characterized by the conditional probability distribution 
function $P(\delg\vert\delm)$.
The constant linear biasing factor $b$ is replaced by a mean {\it biasing
function},    
\be
\coav\equiv b(\delm)\,\delm ,
\label{eq:cond_def}
\ee
which can in principle take a wide range of functional forms,
restricted by definition to have $\av{\delg}=0$ and $\coav\geq -1$ 
for any $\delm$.
The stochasticity is expressed by the higher moments about this mean, 
such as the conditional variance
\be
\sigb ^2(\del) \equiv \av{\epsilon^2 |\delm} /\sigma^2 ,
\quad \epsilon \equiv \delg-\av{\delg|\delm} \ ,
\ee
scaled for convenience by the variance of mass fluctuations, 
$\sigma^2\equiv\av{\delm^2}$.
To second order, the biasing function $b(\del)$ can be characterized 
by two parameters: the moments $\bh$ and $\bt$, 
\be
\bh \equiv\ \av{b(\del)\, \del^2} /\sig^2
\quad {\rm and} \quad
\bt^2 \equiv\ \av{b^2(\del)\, \del^2} /\sig^2 \ .
\ee
The parameter $\bh$ is the natural extension of the linear biasing
parameter, measuring the slope of the linear regression of $\delg$ on
$\delm$, and $\bt/\bh$ is a useful measure of non-linearity.
The stochasticity is characterized independently by a third parameter,
$\sigb ^2 \equiv \av{\epsilon^2}/\sig^2$.
As has been partly explored by Dekel \& Lahav (1999),
these parameters should enter any nonlinear analysis aimed at extracting 
the cosmological density parameter $\Omega$ from a galaxy redshift survey, 
and are therefore important to measure.

In this paper we propose a simple method to measure the biasing function
$b(\delta)$ and the associated parameters $\bh$ and $\bt$ from observed
data that are either already available, such as the PSC$z$ redshift 
survey (Saunders \etal 2000), or that will soon become available, 
such as the redshift surveys of 2dF (Colless 1999) and SDSS 
(\eg, Loveday \etal 1998)
and high-redshift surveys such as DEEP (Davis \& Faber 1999).
Alternative methods have been proposed to measure the 
biasing function, using the cumulant correlators of the observed 
distribution of galaxies in redshift surveys (Szapudi 1998)
or their bispectrum (Matarrese, Verde, Heavens 1997, Verde \etal 1998).

We first show in \se{CDF}, using halos and galaxies in $N$-body simulations,
that the difference between the cumulative
distribution functions (CDFs) of galaxies and mass can be straightforwardly
translated into $\coav$ despite the scatter in the biasing scheme.
Then, in \se{rob}, we demonstrate that for our purpose,
$\cm(\delm)$ is insensitive to the cosmological model and can be 
approximated robustly by a cumulative log-normal distribution.
This means that we do not need to observe $\cm(\delta)$, which is hard to
do; we only need to measure $\cg(\delg)$ and, independently,
the rms value $\sigm$ of the mass fluctuations on the same scale. 
In \se{redshift}, 
we slightly modify the method to account for redshift-space distortions,
and use mock galaxy catalogs from N-body simulations to evaluate the
associated errors.
Finally, in \se{errors}, we estimate the errors due to the sparse
sampling and finite volume.   
The method and its applications to existing and future data are
discussed in \se{conc}.

\section{BIASING FUNCTION FROM DISTRIBUTION FUNCTIONS}
\label{sec:CDF}

Let $\cg(\delg)$ and $\cm(\delm)$ be the cumulative distribution
functions of the density fluctuations of galaxies and mass respectively
(at a given smoothing window).
Had the biasing relation been deterministic and monotonic, it could have
been determined straightforwardly from the difference between these CDFs
at given percentiles,
\be
\delg(\del) = \cg^{-1} [\cm(\del)] \ , 
\label{eq:cc}
\ee
where $\cg^{-1}$ is the inverse function of $\cg$.$\,$\footnote{A similar 
relation has been used by Narayanan \& Weinberg (1998) for ``debiasing" 
the galaxy density field for the purpose of dynamical reconstruction.}
In the presence of scatter in the biasing scheme, strict monotonicity is
violated, but it is possible that $\cg^{-1} [\cm(\del)]$ is still
a good approximation for $\coav$, 
as long as the latter is monotonic.\footnote{The absence of spiral
galaxies in the centers of rich clusters may result in a non-monotonic
biasing function for this type of galaxies at small smoothing scales,
as hinted in Blanton \etal (1999). However, using the simulations
described in this section, 
Somerville \etal (2000) do not find non-monotonicity for late 
type galaxies at $8\hmpc$ smoothing, as used in \Figu{cc_rel}
below.}
The validity of this approximation is addressed in the present section.

We use two cosmological $N$-body simulations in which 
both halos and galaxies were identified (Kauffmann \etal 1999).
The cosmological models are $\tau$CDM (with $\om=1$ and $h=0.5$)
and $\Lambda$CDM (with $\om=0.3$, $\ol=0.7$ and $h=0.7$).
$N=256^3$ particles were simulated in a periodic box of comoving size $85$ and 
$141\hmpc$ respectively (corresponding to a mass resolution of
$1.0\cdot 10^{10}h^{-1}M_{\odot}$ and $1.4\cdot 10^{10}h^{-1}M_{\odot}$).
The simulations were run
using a parallel adaptive P$^3$M code kindly made available by
the Virgo Consortium (see Jenkins \etal 1998) as part of the
``GIF'' collaboration between the HU Jerusalem and the MPA Munich.
The present epoch is defined by a linear rms density fluctuation in
a top-hat sphere of radius $8\hmpc$ of
$\sigma_8=0.6$ in the $\tau$CDM simulation and $\sigma_8=0.9$ in the
$\Lambda$CDM simulation. 
Dark-matter halos were identified at densely sampled time steps using a
friends-of-friends algorithm.
Galaxies were identified inside these halos by applying in retrospect
semi-analytic models (SAMs) of galaxy formation (Kauffmann \etal 1999).
The SAMs simulate the important physical processes of galaxy formation
such as gas cooling, star formation and supernovae feedback. 
At different times in the redshift range 0 to 3, 
we select halos by mass and galaxies by luminosity or type.
We then compute density fields by applying top-hat smoothing with radii
in the range $5-15\hmpc$. We report detailed results for the case
of $8\hmpc$ smoothing, and refer to the scale dependence in several places.

\begin{figure} [t!]
\vspace{11.0truecm}
{\includegraphics{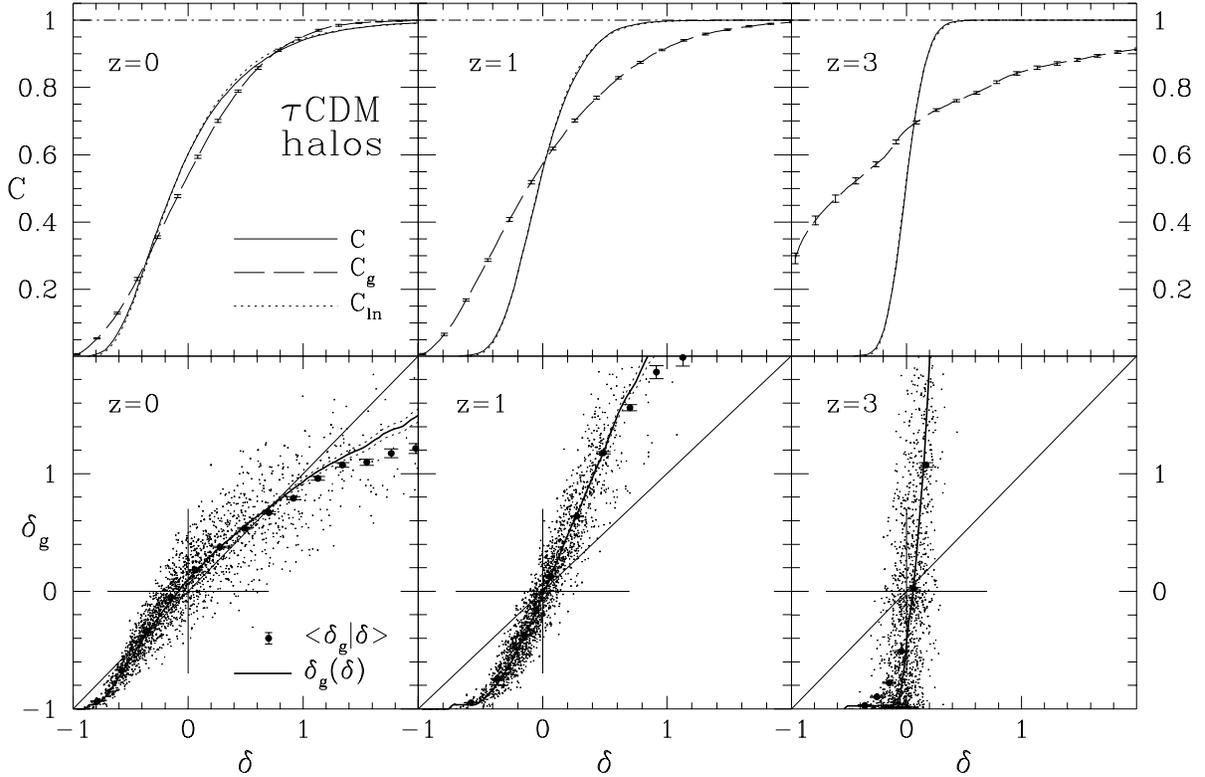}}
\caption{\capt
CDFs and the biasing function at different redshifts
for $\tau$CDM halos with $M> 10^{12}h^{-1}\msolar$ and TH8 smoothing. 
{\it Top panels}: the matter $\cm(\del)$ (solid) and the halo $\ch(\delh)$ 
(dashed). Also shown is a log-normal distribution (dotted), largely hidden 
behind the exact mass distribution. 
{\it Bottom Panels}: $\delh(\vx)$ versus $\del(\vx)$ at grid points
within the simulation box.
The true mean biasing function $\coavh$ is marked by the filled circles
with error bars. Shown in comparison (solid line) is the approximation
obtained by \equ{cc} from the CDFs and the corresponding $1\sigma$ error 
range (dotted).
}
\label{fig:cc_h}
\end{figure}

The figures of this section illustrate the success of the
approximation, \equ{cc}, in several different cases based on the
$\tau$CDM simulation, with top-hat smoothing of radius $8\hmpc$
(hereafter TH8, or TH$X$ for radius $X\hmpc$), 
and at different redshifts.
\Figu{cc_h} refers to halos of mass $> 10^{12}h^{-1}\msolar$ 
($>100$ particles).
On the top we show the cumulative distributions of halos and underlying 
mass fluctuations, $\ch(\delh)$ and $\cm(\del)$
(our notation does not distinguish between halos and galaxies).
The errors in $\ch$ are computed from 20 bootstrap simulations 
of the halo field. The errors in $\cm$, estimated in the same way, 
are smaller by an order of magnitude and are therefore not shown. 
The bottom panels show a point-by-point comparison of the TH8 fields 
of $\delh(\vx)$ and $\del(\vx)$ at points randomly chosen (1:8)
from a uniform grid of spacing $2.64\hmpc$ within the simulation box.
The true mean biasing function $\coavh$ 
is marked by the filled circles with attached error bars. 
It is computed by a local linear regression of $\delh$ on $\delta$ within 
each bin of $\delta$, adopting the value of the fitted line at the 
center of the bin (only every other bin is shown).
Shown in comparison (solid line) is the approximation for $\coavh$ obtained by 
\equ{cc} from the CDFs, and the corresponding $1\sigma$ error range based
on the bootstrap realizations (dotted lines).

As can be seen in \Figu{cc_h}, the approximation is excellent over most
of the $\delta$ range --- the deviation at $z$=0 is within the
$1\sigma$ errors 
up to $\delta \sim 1.4$ (corresponding to $\sim 97\%$ of the volume). 
Systematic deviations show up at higher $\delta$ values, where the 
scatter becomes larger and the mean biasing function flatter,
making the deviations from monotonicity larger.
In order to quantify the quality of the approximation, we average  
the residuals (scaled by $\sigg$):
\be
\Delta = {{1}\over{N_{\rm bins} \sigg^2}}
\sum_{\delm-{\rm bins}}^{N_{\rm bins}}  {[\delg(\delm) - \coav]^2} \ ,
\label{eq:delta}
\ee
where $\delg(\delm)$ is obtained via \equ{cc}.
We exclude the poorly recovered high-density tail by
limiting the summation to those $N_{\rm bins}$ bins of $\delm$ for which
$\cm(\delm)<0.99$ and $\cg(\delg)<0.99$.
The values of $\Delta$ in the various cases studied, including
halos and galaxies in $\tau$CDM and $\Lambda$CDM at different redshifts,
are listed in Table~1.
For example, for the halos shown in \Figu{cc_h} at $z=0$
we obtain $\Delta = 0.08$, indicating that the typical error in
the approximation $\delh(\delm)$ is small compared to the actual scatter 
$\sigh$ in the halo density field. 

A complementary approach for quantifying the quality of the approximation
is by testing how well it recovers the values of the
moments of the biasing function, $\bh$ and $\bt$. In Table~1 we present 
the values of these moments for the different cases, as computed directly 
from the simulation and as approximated by $\delg(\delm)$  
(denoted by a subscript ``a"). 
These biasing parameters are computed based on 99.9\% of the volume,
excluding the
very highest density peaks, where the error is 
large
(The only exception is at $z$=3, where we use only 99\% of the volume
because the errors are even larger).
For the halos shown in \Figu{cc_h} at $z=0$,
we see that $\bh$ and $\bt$ are recovered with errors of  
1\% and 3\% respectively.

\begin{figure} [b!]
\vspace{11.0truecm}
{\includegraphics{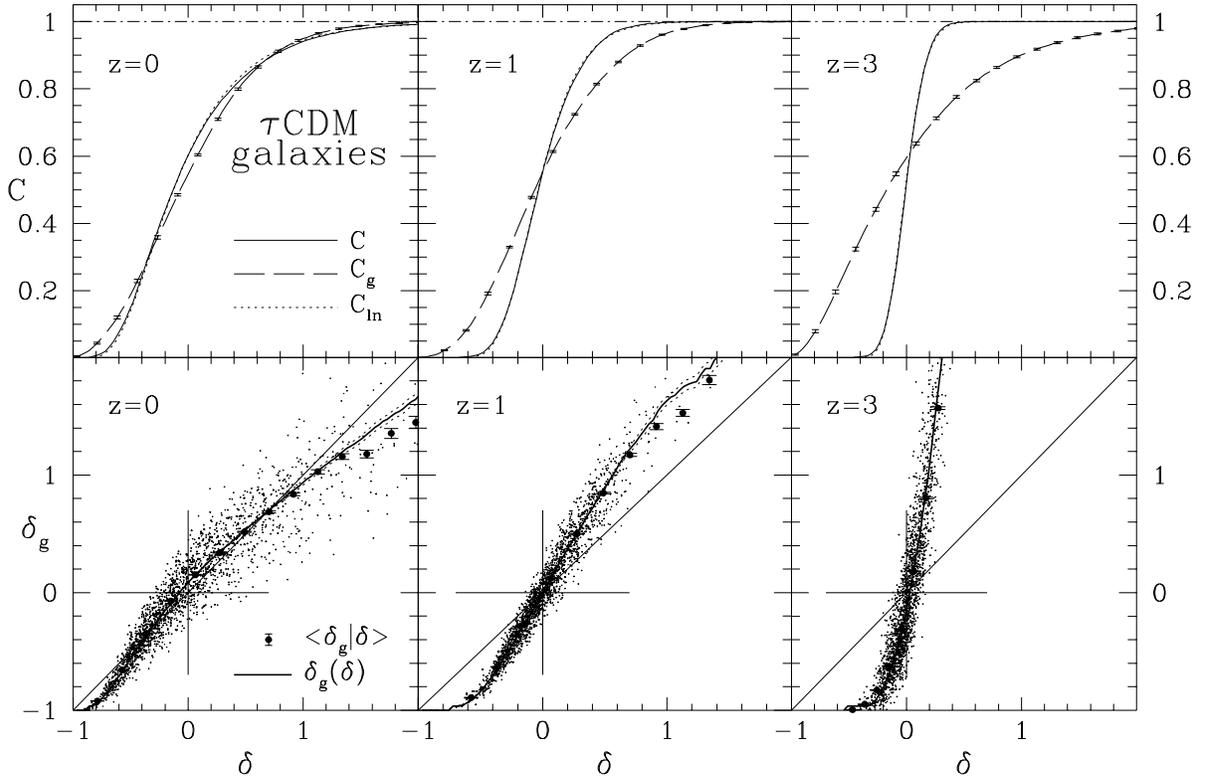}}
\caption{\capt
Same as \Figu{cc_h}, but for bright galaxies of $M_{\rm B}< -21$ rather than
massive halos.
}
\label{fig:cc_g}
\end{figure}

The middle panels of \Figu{cc_h} refer to $z=1$, where $\bh\simeq 2.2$.
The approximation of \equ{cc} holds well in this case  
up to $\delm\sim 0.7$, which corresponds to 
$\sim 98\%$ of the volume. 
The approximation remains good despite the large scatter (compared
to the $z=0$ case) because the steepness of the biasing function
helps maintaining reasonable monotonicity. The goodness of the
recovery of the biasing function, with $\Delta = 0.07$,
is similar to the $z=0$ case. The parameters $\bh$
and $\bt$ are recovered with an accuracy of $\sim 5\%$ (Table~1).
The right panels of \Figu{cc_h} demonstrate that the approximation is 
valid even at $z$=3, where the biasing is extremely strong, $\bh\simeq 6.6$. 
The recovery of the biasing function is still good, $\Delta = 0.20$, 
and its moments are approximated to within $\sim 2\%$.
 
The halo biasing function in the $\Lambda$CDM cosmology is recovered,
in general, with similar success, as can be seen in the top part 
of Table~1. 
Note that in this case the recovery actually improves at higher redshift. 
This reflects the fact that in $\Lambda$CDM the halo biasing scatter becomes 
smaller at higher redshift (see Somerville \etal 2000, Fig. 17).
It results from the 
smaller shot noise due to the higher abundance of high-redshift halos  
in $\Lambda$CDM compared to $\tau$CDM.


\Figu{cc_g} is analogous to \Figu{cc_h}, but now for bright galaxies of 
$M_{\rm B}-5\log h < -19.5$. The recovery is again
quantified in Table~1; it is quite similar to the case of halos.
The typical error is $\Delta\leq 0.08$,
and the biasing parameters are recovered with an error of a couple to
a few percent.

The performance of our method has been tested for smoothing scales
in the range  $5-15\hmpc$. 
For the $\tau$CDM model, we find that the quality of the approximation
is practically independent of scale throughout this range; 
the relative error in the biasing parameters is at the level of a few 
percent, and $\Delta$ is in the range 0.1 to 0.2, rather similar to the 
values quoted in Table~1 for TH8 smoothing. 
On the other hand, for $\Lambda$CDM we do find that the performance
improves with increasing smoothing scale.
With TH15 at $z=0$, for halos (or galaxies),
the errors in the biasing parameters reduce to below 3\% (1\%),
and $\Delta=0.07$ (0.04), while for TH5 smoothing these errors
are about 4 times larger.
This difference between the two models can be attributed to a 
difference in the scale dependence of the biasing scatter 
(Somerville \etal 2000, Figure 16), which translates to an error
in our method via increased deviations from monotonicity.

Before we proceed with the biasing relative to the underlying mass,
we note that the {\it relative~} biasing function of two
different galaxy types, $\av{\delgtwo | \delgone}$, can be directly 
observable from a redshift survey.
Again, for a deterministic and monotonic biasing process one has
\be
\delgtwo (\delgone ) = \cgtwo^{-1}[\cgone(\delgone )] \ ,
\label{eq:cc12}
\ee
and when biasing scatter is present, the question is to what extent
\equ{cc12} provides a valid approximation for the true relative
biasing function.

\begin{figure} [t!]
\vspace{9.truecm}
{\includegraphics{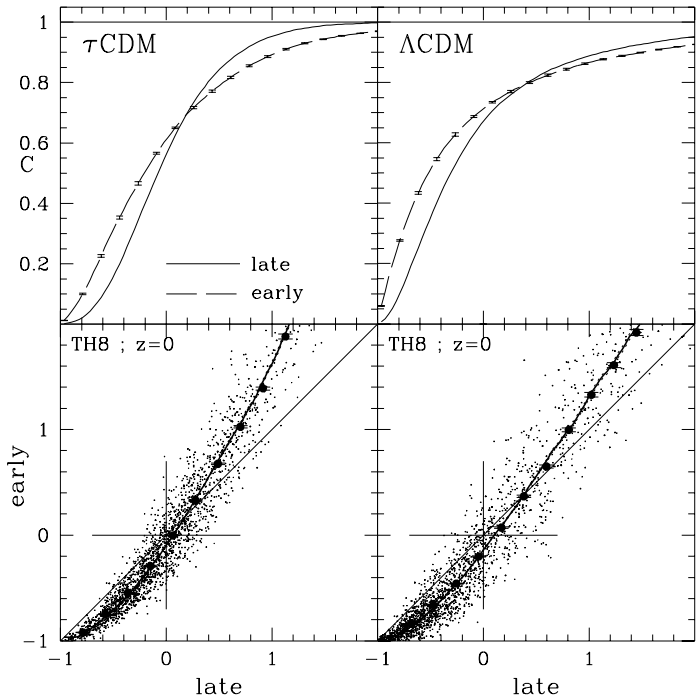}}
\caption{\capt
The relative biasing of early versus late type galaxies, at $z$=0,
for $\tau$CDM (right panels) and $\Lambda$CDM (left panels).
The symbols are as in \Figu{cc_h}.  
}
\label{fig:cc_rel}
\end{figure}

\Figu{cc_rel} shows the
relative biasing function of ``early'' and ``late'' type galaxies
in the two cosmological models, at $z=0$ and with TH8 smoothing.
These galaxy types are distinguished in the SAM $N$-body simulations
according to the ratio of bulge to total luminosity in the V band being
larger or smaller than 0.4 respectively. 
The large scatter in the relative biasing, due to errors in the two 
density fields, is reduced by including all the galaxies, without applying 
a luminosity cut.

As can be seen in the last three columns of Table~1, the quality of the 
recovery of the relative biasing function is not as good as in
the case of the absolute biasing of galaxies or halos.
The values of $\Delta$ range from 0.2 to 0.56, compared to
0.08 to 0.16 in the former cases.  This is expected, because in the
case of relative 
biasing the two density fields contribute to the stochasticity
or deviation from monotonicity 
(see also the important role of sampling 
errors in the recovery of the biasing function, \S4.2).
The moments of the relative biasing function are recovered 
with better than 15\% accuracy at $z\leq 1$, and to $\sim$25\%
accuracy at $z=3$, in both cosmologies. In calculating the moments,
unlike in \Figu{cc_rel}, a luminosity cut has been applied:
$M_{\rm B}-5\log h < -19.5$, 
and 99\% of the volume was used.
The fact that the $\Delta$ values are still significantly smaller than
unity and the errors in the biasing parameters are not larger than 25\%
indicate that our method is capable of yielding meaningful estimates
of the relative biasing function.
In both cosmologies, the relative biasing is almost scale independent
in the range 5 -- 15 $\hmpc$, as is the quality of the reconstruction.

\section{THE MASS CDF: ROBUST AND LOGNORMAL}
\label{sec:rob}

Large redshift surveys provide a rich body of data for mapping 
the galaxy density field in extended regions of space and computing 
its CDF with adequate accuracy. However, direct mapping of the {\it mass}
density field is much harder. For example, POTENT reconstruction from 
peculiar velocities 
(Dekel, Bertschinger \& Faber 1990; Dekel \etal 1999; Dekel 2000) 
yields the mass distribution in our local cosmological neighborhood
(even out to $\sim 100\hmpc$), which in principle enables direct 
mapping of the local biasing field. However, the sparse and noisy data 
limit the mass reconstruction to low resolution ($\sim 10\hmpc$)
compared to the volume sampled, which introduces 
large cosmic scatter in the mass CDF. 
New accurate data nearby, based on SBF distances (Tonry \etal 1997)
do enable a promising resolution of a few Mpc (see Dekel 2000), but 
limited to inside the local sphere of radius $\sim 30\hmpc$.

What makes the method proposed here feasible is the fact that the mass
CDF is only weakly sensitive to variations in the cosmological scenario 
within the range of models that are currently considered as viable
models for the formation of large-scale structure (\eg, Primack 1998,
Bahcall \etal 1999).
It has been proposed that the mass PDF can be well approximated by a
log-normal distribution in $\rho/\bar\rho=1+\delta$
(\eg, Coles \& Jones 1991; Kofman \etal 1994),
and it has since been argued that this approximation becomes poor 
for certain power spectra and at the tails of the distribution
(Bernardeau 1994; Bernardeau \& Kofman 1995).
In this section, we investigate the robustness of $\cm(\del)$ for our
purpose here, namely, in comparison with the typical difference between 
the CDFs of galaxies and mass (\ie, the mean biasing function) which we are
trying to approximate.

We use for this purpose a suite of $N$-body simulations of six different 
cosmological models. In addition to the two high-resolution simulations 
of $\tau$CDM and $\Lambda$CDM used in the previous section,   
we have simulated three random realizations of each of the three following
models (all using a Hubble constant of $h=0.5$):  
standard CDM (SCDM; $\om=1$ with spectral index $n=1$), 
an extreme open CDM (OCDM; $\om=0.2$, $n=1$),
and an extreme tilted CDM (TCDM; $\om=1$, $n=0.6$).
These simulations were run by Ganon \etal (2000, in preparation) using
a PM code (by Bertschinger \& Gelb 1991), 
with $128^3$ particles in a $256\hmpc$ box.
The present epoch is defined in these simulations by a linear fluctuation 
amplitude of $\sigma_8=1.0$.
A similar simulation was run using a constrained realization (CR) of the
local universe based on the galaxy density in the IRAS 1.2Jy redshift survey 
under the assumption of no biasing (Kolatt \etal 1996), 
with $\om=1$ and the present defined in this case by $\sigma_8=0.7$.

\begin{figure} [t!]
\vspace{7.7truecm}
{\includegraphics{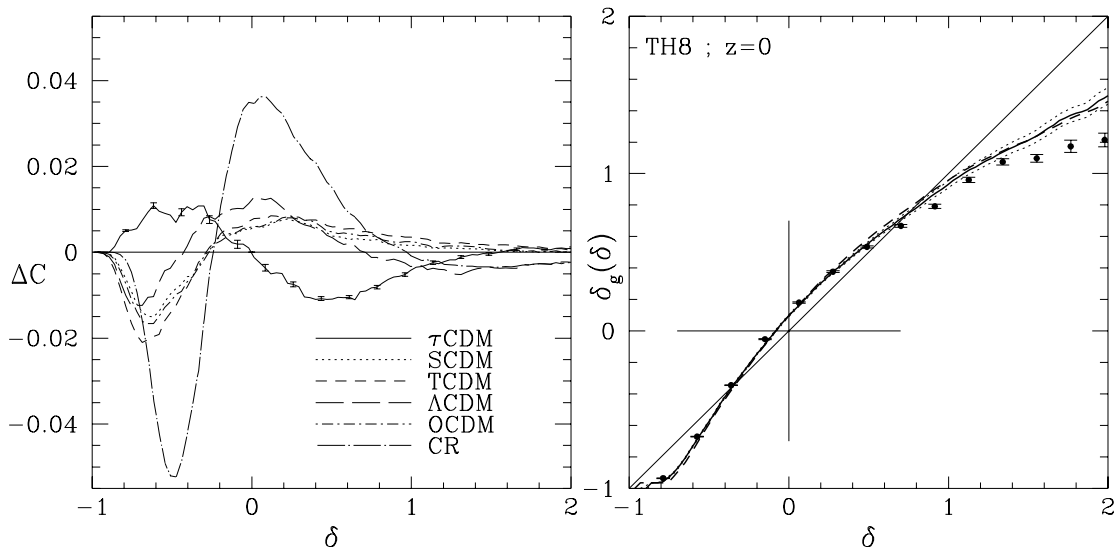}}
\caption{\capt
Robustness of the mass CDF to cosmological models.
{\it Left}: The deviation $\Delta C$ of the CDFs from a cumulative 
log-normal distribution, for various CDM cosmologies at $z=0$: 
$\tau$CDM (solid); 
$\Lambda$CDM (long-dashed); OCDM (dot-dashed); TCDM (dashed); 
SCDM (dotted); and CR (dot-long-dashed). 
{\it Right}: The approximation $\delh(\delm)$ based on the exact
$\cm({\delm})$ (solid curve, with dotted lines marking 1-$\sigma$ errors), 
versus the one based on the approximation $\cm({\delm})=\cln(\delm)$ instead 
(dashed curve). They lie almost on top of each other.
The true mean biasing function $\coavh$ is shown for comparison
(points with error bars).
All are for halos with $M>10^{12}\hmsun$ in the $\tau$CDM simulation.
}
\label{fig:lognormal}
\end{figure}

\Figu{lognormal} (left) shows for the different models the deviations 
$\Delta \cm(\delta)$ of the mass CDFs, smoothed TH8, from a cumulative
log-normal distribution with the same $\sigma$. 
The log-normal probability density is 
\def\trho{\tilde\rho}
\be
\pln(\delta) = {1\over \trho} {1\over \sqrt{2\pi} s}
                  \ \exp \left[-{ (\ln\trho-m)^2 \over 2s^2} \right] \ ,
\label{eq:logn}
\ee
where
\be
\trho=1+\delta \ , \quad
m=-0.5 \ln(1+\sigma^2) \ , \quad 
s^2=\ln(1+\sigma^2)  \quad {\rm and} \quad
\sigma^2 = \av{\delta^2} \ .
\ee
The cumulative log-normal distribution is obtained by integration,
\be
\cln(\delta) = {\rm erf} \left[ {\ln\trho - m \over s} \right] \ ,
\label{eq:cln}
\ee
where
\be
{\rm erf}(x) \equiv {1\over \sqrt{2\pi}}  \int_{-\infty}^x 
                    e^{-t^2/2} \dd t \ .
\ee
For the cases of OCDM, TCDM and SCDM, the CDF 
is obtained from the three simulations of each model put together.
The errors are similar in the different cases; 
we therefore plot representative error bars only for the $\tau$CDM case. 
 
In all the realizations that had random Gaussian initial conditions, 
the deviation from lognormality is less than 2\%. The constrained 
realization shows somewhat larger deviations, but even in this case they
never exceed 5\%. These deviations are indeed smaller than the
typical differences between $\ch(\delm)$ and $\cm(\delm)$, 
which are on the order of 10\% (see \Figu{cc_h}).

In order to evaluate how important the contribution of $\Delta C$ is 
to the error in the recovery of $\coavh$, we compare in the right 
panel of \Figu{lognormal} the true $\coavh$ in the $\tau$CDM simulation
with two approximations $\delh(\delm)$ based on \equ{cc}, 
one using the true matter CDF and the other replacing it with a 
cumulative log-normal distribution of the same $\sigm$. The results of
the two approximations are very similar; the differences between them
seem to be much smaller than the differences between each of them and the
true biasing function $\coavh$.
We can conclude that for the purpose of recovering the biasing function, 
for the range of 
Gaussian 
cosmological models considered, $\cln$ is a good approximation for $\cm$. 

The proximity of $\cm$ and $\cln$ could have been alternatively evaluated by
the Kolmogorov-Smirnov (KS) statistic, $D={\rm max}\{\vert\Delta C\vert\}$.
For computing the KS significance $q(D)$, we estimate the effective
number of ``independent" points by $\neff=\volbox/\vwin$, where $\volbox$ 
is the volume of the simulation box and $\vwin$ is the effective volume of
the smoothing window. A value of $q\simeq 1$ ($D\ll 1$) corresponds to a 
good match, and $q\ll 1$ ($D \simeq 1$) to a poor match.
For our $\tau$CDM simulation, with TH8 smoothing at $z=0$ and 1,
we obtain $D\simeq 0.01$ and $q>0.9999$, confirming that $\cln$ is a
good fit. However, for the larger SCDM and OCDM simulations, although
$D$ is still only $\simeq 0.015$, the corresponding $q$ values are at the
level of only a few percent. 
For TCDM and CR, where $D$ is 0.016 and 0.052 respectively, the values
of $q$ drop to the level of a fraction of a percent, and the discrepancy
seems large.  This KS test indicates that the log-normal approximation 
is not always perfect for general purpose, as has been argued in 
the literature.  However, our direct tests reported above demonstrate
that the use of the log-normal approximation is adequate 
for the recovery of the mean biasing function in all these cases.

We comment in passing that
while the mass CDF is well approximated for our purpose by a log-normal 
distribution, the shape of the halo (or galaxy) CDF is usually far from 
a log-normal shape. This is implied by \equ{cc}, from which it follows that
$\cg (\delg )= \cm [\delg ^{-1}(\delg)]$.
One does not expect to recover a log-normal distribution from a
general functional form for $\delg ^{-1}$.
In particular, the linear biasing model, which seems to be an acceptable
approximation in some cases with large smoothing (\eg, IRAS 1.2Jy galaxies 
at 12$\hmpc$ Gaussian smoothing; Sigad \etal 1998), 
leads to a $\cg (\delg)$ that is far from log-normal.
Trying to evaluate the difference between $\ch$ and a log-normal
distribution using the KS statistic, we obtain for the halos in the $\tau$CDM
simulation, with TH8 smoothing, both at $z=0$ and 1,
$D\simeq 0.08$ and $q\simeq 0.05$, namely a poor fit compared to the
$q\simeq 1$ of $\cm$ vs $\cln$.
Similar conclusions are valid for galaxies.

Our method for measuring the nonlinear biasing function requires 
an assumed value of $\sigm$. Since $\sigm$ is known only to a limited
accuracy (\se{conc}), we should check the robustness of our results to
errors in $\sigm$. We repeated the reconstruction described in \se{CDF},
both for halos and for galaxies,
with perturbed values of $\sigm$ in a range $\pm 20\%$ about the
true value of the simulation.
Not surprisingly, we find that the analog of the linear biasing
parameter, $\bh$, varies roughly in proportion to $\sigm^{-1}$. 
We also find that $\bt$ varies in a similar way, such that the
ratio $\bt/\bh$, which is the natural measure of nonlinear biasing
(Dekel \& Lahav 1999), is a very weak function of $\sigm$, 
roughly $\bt/\bh \propto \sigm^{0.15}$. This test indicates that
our method provides a robust measure of the nonlinearity in the biasing
scheme, that is to a large extent decoupled from the uncertainty in the
linear biasing parameter.

\section{REDSHIFT DISTORTIONS}
\label{sec:redshift}

The densities as measured in redshift space (z-space)
are in general different from the real-space (r-space)
densities addressed so far, because the radial peculiar velocities 
distort the volume elements along the lines of sight.
One approach to deal with redshift distortions is to start by recovering 
the full galaxy density field in r-space, 
using the linear or a mildly-nonlinear approximation to gravitational
instability (e.g., Yahil \etal 1991;  Strauss \etal 1992;
Fisher \etal 1995; Sigad \etal 1998), 
and then compute the biasing function in r-space as outlined above.
The accuracy of such a procedure would be limited by the approximation
used for nonlinear gravity.
Another difficulty with this approach is that it requires one to assume a
priori a specific 
biasing scheme, already in the force calculation that enters the 
transformation from z-space to r-space, while this biasing scheme is 
the very unknown we are after; this would
require a nontrivial iterative procedure.  

The alternative we propose here is to actually use the z-space
CDF, $\cgz(\delgz)$, as provided directly from counts in cells of 
galaxies in a redshift survey.
If the redshift distortions affect the densities of galaxies and mass
in a similar way, then one may expect the biasing function in z-space to 
be similar to the one in real space, 
\be
\av{\delgz | \delmz\!=\!\delm} = \av{\delg | \delm} \ .
\label{eq:bf_zr}
\ee
If we only had a robust functional form for the mass CDF in
z-space, $\cmz(\delmz)$, then we could compute the desired biasing
function all in z-space, using \equ{cc} but with the analogous z-space
quantities. We thus need to test the validity of \equ{bf_zr}, and come up
with a useful approximation for $\cmz(\delmz)$.

\begin{figure} [t!]
\vspace{8.0truecm}
{\includegraphics{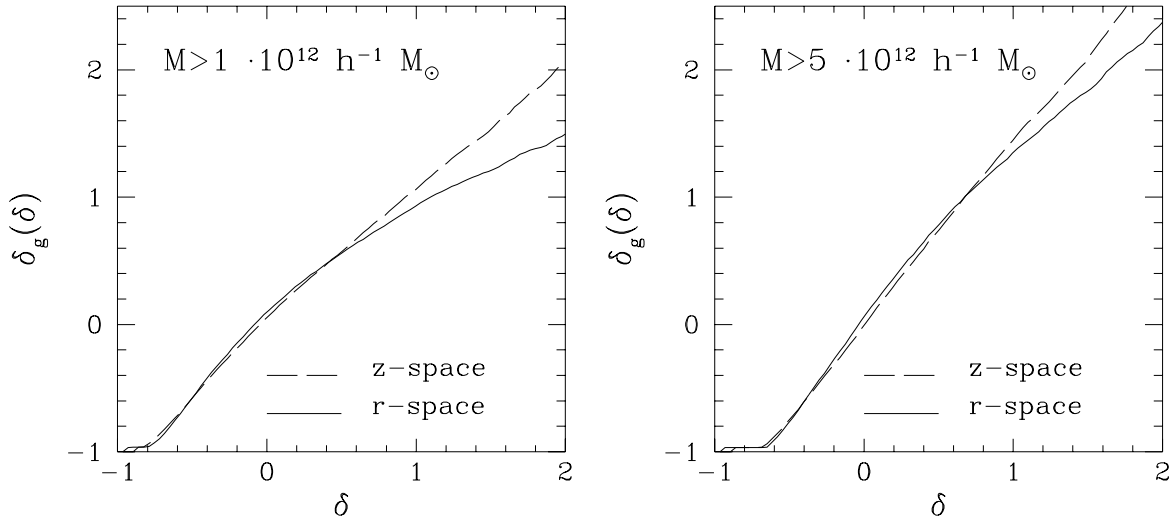}}
\caption{\capt
Biasing functions in z-space (dashed) versus r-space (solid).
The biasing functions are derived from the corresponding TH8 CDFs of
halos and mass in the $\tau$CDM simulation at $z=0$.
Shown are halos of $M> 10^{12}h^{-1}\msolar$ (left) and
$M> 5\cdot 10^{12}h^{-1}\msolar$ (right).
}
\label{fig:bias_z1}
\end{figure}

\Figu{bias_z1} illustrates the accuracy of \equ{bf_zr}.
It compares the biasing functions in z-space and r-space,
as derived via \equ{cc} and its z-space analog from the corresponding 
CDFs of halos and mass in the $\tau$CDM simulation with TH8 smoothing.
The two curves are remarkably similar for $\delm < 0.6-0.8$, 
roughly out to the 1-sigma rms fluctuation value. 
This is roughly the range where the biasing scatter is reasonably small
and our basic method is applicable (\se{CDF}, \Figu{cc_h}).
The curves deviate gradually as $\delm$ increases, partly due to stronger
``fingers of god'' effects at high densities. The deviation is somewhat
weaker for larger-mass halos
(perhaps due to a lower velocity dispersion for more massive objects
as a result of dynamical friction).

The direction of the deviation from \equ{bf_zr}, as seen in \Figu{bias_z1},
can be obtained by applying linear theory of redshift distortions
to the case of linear biasing in r-space, $\delg = b \delm$.
In linear theory, the density fluctuations in r-space and z-space 
are related via  $\delmz=\delm [1+f(\om)\mu^2]$, where
$f(\om) \simeq \om^{0.6}$ 
(with a negligible dependence on $\ol$,
see Lahav \etal 1991) and $\mu$ is the
cosine of the angle between the galaxy velocity vector and the line of sight.
If the galaxies obey the continuity equation, then $\delgz-\delg=\delmz-\delm$,
which implies the following biasing relation in z-space:
\be
\delgz = {b+f(\om)\mu^2 \over 1+f(\om)\mu^2} \, \delmz \ .
\ee
Averaging over all possible directions and assuming
$\om=1$, we find that the linear biasing parameter in z-space 
is predicted to be $b_z=(3b+2)/5$ for the case shown in \Figu{bias_z1}.
This indicates that the linear biasing
parameter tends to be closer to unity in z-space than in r-space.
Based on our empirical tests of \equ{bf_zr}, we learn that the
nonlinear effects (of biasing and gravity) conspire to make \equ{bf_zr}
a better approximation than implied by the linear approximation.

Note that while the results of \Figu{bias_z1} based on our high-resolution 
$\tau$CDM simulation are quite accurate in the way they treat halos, they may
suffer from significant cosmic variance due to the relatively small 
volume sampled, where the presence (or absence) of a few ``fingers of god" 
could strongly affect the biasing function in the high-$\delm$ regime.
To test the validity of \equ{bf_zr} with reduced cosmic variance,
we appeal to yet another set of N-body simulations (by Cole \etal 1997)
which cover a much larger volume but with lower resolution.
These simulations followed the evolution of $N=192^3$ particles in 
a periodic box of comoving side $L=345.6 \hmpc$ 
using an Adaptive P$^3$M code. The cosmological models are $\Lambda$CDM 
($\om=0.3$, $\ol=0.7$, $h=0.65$, cluster-normalized to $\sigma_8=1.05$)
and $\tau$CDM ($\Omega=1$, $h=0.25$,  cluster-normalized to $\sigma_8=0.55$).
Nine mock catalogs were extracted from each of the parent simulations,
each containing $\sim 5 \cdot 10^5$ particles in a box of $L=200 \hmpc$. 
The partial overlap between the catalog volumes is thus about $50\%$.
The central ``observer" was chosen to mimic certain properties of the 
Local Group environment (see Branchini \etal 1999).
Since the resolution of these large simulations is inadequate for a
detailed halo identification based on many simulated particles in each halo, 
we identify individual particles as galaxies using a Monte-Carlo procedure
in which the galaxies are chosen to make a random realization of an 
assumed nonlinear biasing function. Here we adopt the biasing 
function proposed by Dekel \& Lahav (1999) to fit the simulated 
results of Somerville \etal (2000):
\be
\delg(\delta)= \left\{ \brr{ll}
        (1 + b_0)(1 + \delta)^{b_{\rm neg}} -1 & \delta \le 0 \\
         b_{\rm pos}\delta  +  b_0 & \delta > 0
        \err \right\} ,
\label{eq:nlbias}
\end{equation}
with $b_{\rm neg}=2$ and $b_{\rm pos}=1$.
The mass density field is obtained with a Gaussian smoothing of
radius $5\hmpc$ at the points of a $128^3$ cubic grid inside a box
of size $200\hmpc$.
Galaxy densities are obtained at the grid points based on
\equ{nlbias}, and then interpolated to the galaxy positions as defined
by the selected particles.
Given the appropriate probability distributions $P(\delta)$,
the value of $b_0$ is determined for each choice of the parameters
$b_{\rm neg}$ and $b_{\rm pos}$ such that $\langle \delg \rangle =0$ 
as required by definition.
We obtain $b_0=0.26$ and $b_0=0.19$ for the models of $\Lambda$CDM
and $\tau$CDM respectively.

\begin{figure}[t!]
\vspace{11.0truecm}
{\includegraphics{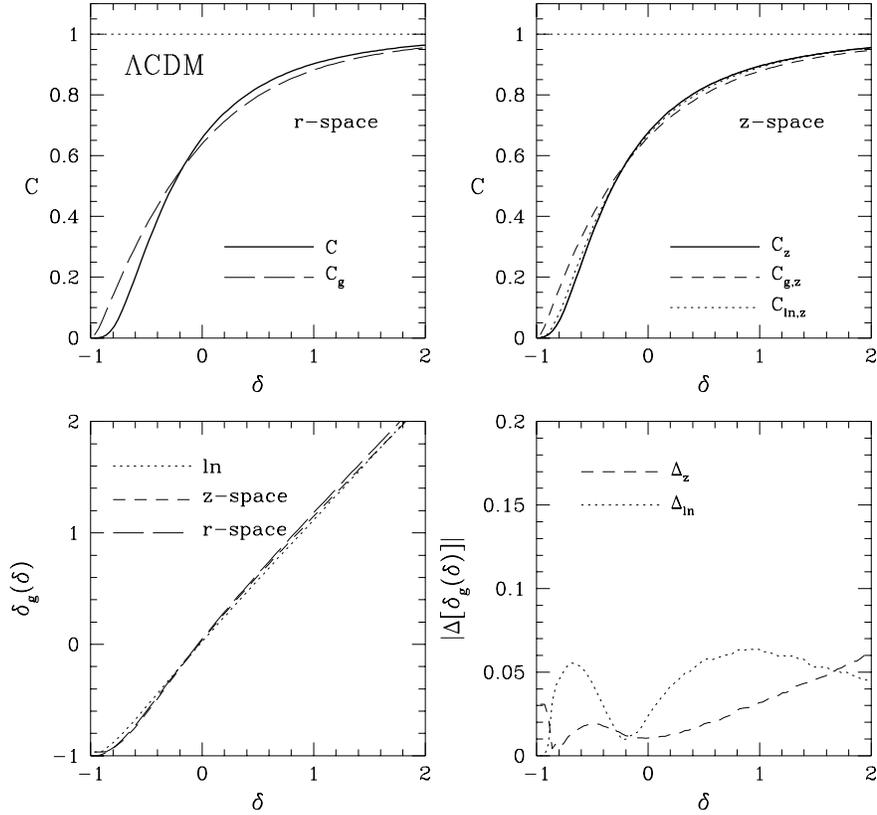}}
\caption{\capt
CDFs and biasing functions in r-space versus z-space, averaged over mock
catalogs that were extracted from the large $\Lambda$CDM simulation,
with TH8 smoothing. 
Top: CDFs in r-space (left) and z-space (right), 
for mass (solid) and for galaxies (dashed). 
Shown in comparison is $\clnz$ with the $\sigmz$ of the matter (dotted).
Bottom left: the biasing functions as derived from
the CDFs in r-space (long-dash) and z-space (short-dash); they are
very similar.  Also shown is the biasing function derived
in z-space assuming $\clnz$ with $\sigmz$ obtained using \equ{sigmz}
(dotted line).
Bottom right: absolute value of the difference between the 
biasing functions: $\delgz(\delmz\!=\!\delm)-\delg(\delm)$ (dashed)
and $\dellnz(\delmz\!=\!\delm)-\delg(\delm)$ (dotted).
}
\label{fig:rz_cole}
\end{figure}

\Figu{rz_cole} compares the CDFs and associated biasing functions 
in r-space and z-space, averaged over nine mock catalog from the large-box
$\Lambda$CDM simulation.
The z-space biasing function is indeed almost indistinguishable from the
r-space one (bottom panels); the differences are typically on the order of a 
couple of percents.
The results for $\tau$CDM are similar.

In order to quantify this difference further, we define a statistic
analogous to 
\equ{delta}:
\be
\Delta = {{1}\over{N_{\rm bins} \sigg^2}}
\sum_{\delm-{\rm bins}}  {[\delgz(\delmz\!=\!\delm) - \delg(\delm)]^2} \ ,
\label{eq:delta2}
\ee
in which the first and second terms are the biasing functions as derived from
the CDFs in z-space and r-space respectively.
The summation is over bins with $\delta < \delta_{\rm max}$, such that
$\approx 99\%$ of the volume is accounted for.
We also compute the two moments of the observed biasing function 
$\bh_{\rm obs}$ and $\bt_{\rm obs}$.
These three quantities, averaged over the mock catalogs, 
are listed in Table~2 (second column). Their deviation from the
``true'' values (Table~2, first column)
is the systematic error.  The quoted errors refer to the 1$\sigma$
scatter about the mean; they represent the random errors. 
The results are listed for the two models, $\Lambda$CDM and $\tau$CDM. 
We conclude that the biasing function and its moments, as computed 
from the z-space CDFs, resemble those computed from the r-space 
CDFs to within 2\%. 
Note that the Monte Carlo procedure
we use to generate mock catalogs artificially reduces the amount of 
clustering and over-smoothes the density fields for dark and luminous 
particles. The net effect is to decrease the biasing 
moments by $\sim 7\%$, relative to the values implied by the biasing
scheme, \equ{nlbias}.
However, this bias does not affect the present 
analysis for which ``true'' values are obtained from the mock catalogs
themselves 
and not from  \equ{nlbias}.

Our next task is to come up with a robust CDF for the mass in z-space.
We try the same log-normal distribution that was found robust for
our purpose in r-space (\se{rob}), but with a proper rms in
z-space, $\sigmz$.
Based on the linear approximation for Gaussian fields in the small-angle
limit (Kaiser 1987), we express $\sigmz$ in terms of $\sigm$
and $\om$ of the cosmological model by:
\be
\sigmz= \left[1 +{2\over3}f(\om) + {1\over5}f^2(\om)\right]^{\onehalf} \,\sigm.
\label{eq:sigmz}
\ee
%
We thus approximate the z-space biasing function by $\dellnz(\delmz)$,
as derived from the z-space CDFs but where the mass CDF is replaced by
a cumulative log-normal distribution function $\clnz$ (\eqd{cln}) 
with standard deviation $\sigmz$ (\eqd{sigmz}).
The resultant biasing function, averaged over the mock catalogs,
is displayed in the bottom panels of \Figu{rz_cole}.
We see that for $\Lambda$CDM the differences between 
$\dellnz(\delmz\!=\!\delm)$ and $\delg(\delm)$ are at at the level 
of a few percent. For $\tau$CDM they are only a bit larger; they
exceed 10\% but only near $\delm \sim 2$, at the tail of the distribution.
The error in the biasing function $\Delta$ defined in analogy to \equ{delta2},
and the biasing moments, are listed in Table~2 (third column, marked
``z-space ln").  
The systematic error $\Delta$ is still well below 2\%, 
but the biasing parameters are systematically underestimated by 
4\% and 7\% in $\Lambda$CDM and $\tau$CDM respectively. 

Overall, it seems that our straightforward method deals with redshift
distortions fairly well, without any a priori assumption about the
biasing scheme.

\section{SAMPLING ERRORS}
\label{sec:errors}

The accuracy of the derivation of the galaxy PDF is limited by two 
observational factors: the finite volume sampled and the mean density 
of galaxies in the sample.\footnote{The additional edge effects can be greatly 
minimized by using a volume-limited sample and a proper choice of cell 
coverage (see Szapudi \& Colombi 1996).}
 
In principle, the limited volume is responsible for cosmic variance
due to the fact that the long-wavelength fluctuations in the real
universe are not
fairly represented in the sampled volume. This is not of major concern
for us here because (a) it is expected to introduce only a random error,
and (b) as long as the biasing is local, the effects of long waves
on the PDFs of galaxies and mass are expected to be correlated, making the
local biasing function representative of the universal function
despite the relatively small sampling volume.

More important is the {\it shot noise} introduced by the combination of
volume and sampling density effects. For a given cell size (or smoothing
length), the error can be divided into the error in the count within each cell
and the error due to the finite number of cells in the sample volume.
These shot-noise sources may introduce both random and systematic errors.
We evaluate them by computing the mean and standard deviation
over a suite of mock catalogs in which we vary either the volume
or the sampling density for a fixed smoothing scale.

\begin{figure}[b!]
\vspace{11.0truecm}
{\includegraphics{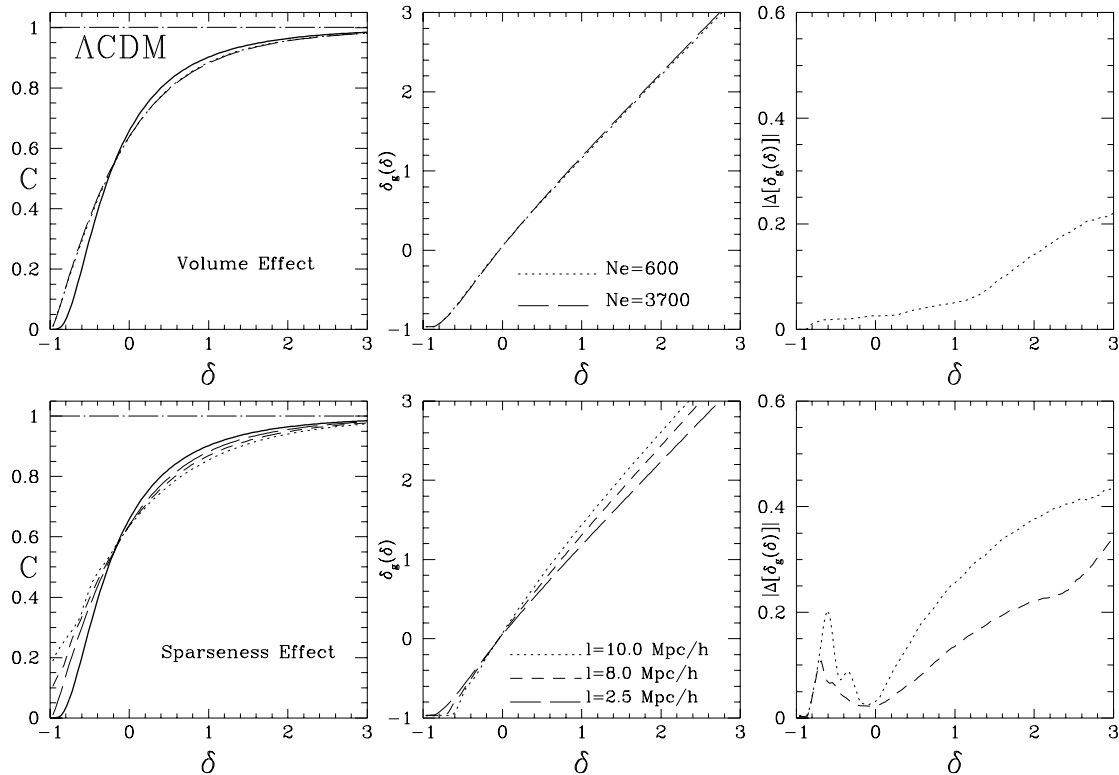}}
\caption{\capt
Sampling errors due to finite volume (top) and sparse sampling
(bottom), for fixed 
TH8 smoothing, estimated from the large $\Lambda$CDM simulation.
Shown are the CDFs in real space (left), the derived biasing function
(middle), and the error in it (right).
The mass CDF is marked by a solid line, and galaxy CDFs by broken lines.
Top: volumes of $\neff=3700$ and 600 are marked by long-dashed and 
dotted lines respectively.
Bottom: samples of galaxy separation $l=2.5, 8,$ and $10 \hmpc$ are marked 
by long-dashed, short-dashed, and dotted lines respectively.
}
\label{fig:errors}
\end{figure}

With TH8 smoothing, our mock catalogs from the large $\Lambda$CDM
simulation contain $\neff \sim 3700$ independent cells.
However, the currently available redshift surveys allow
an analysis in a much smaller volume. For example, a volume-limited
subsample from the PSC$z$ catalog (Saunders \etal 2000),
that is cut at a distance where the average galaxy separation
is $l=8\hmpc$ (i.e., on the order of our smoothing scale),
contains only $\sim 600$ independent cells. 
We therefore estimate the error associated with
reducing the sampled volume such that $\neff\sim 600$ in each mock catalog.
We select from the simulation 9 such non-overlapping sub-volumes,
while keeping the sampling density and smoothing length fixed.
The results for $\Lambda$CDM, averaged over the mock catalogs, are
shown in the upper  
panels of \Figu{errors}, and the results for the two cosmological models are
summarized in Table~2 (column 4).
We find no significant systematic errors due to the
volume effect in a sample like PSC$z$ and with $\sim 8\hmpc$ smoothing
(except in the very high-$\delm$ tail for $\tau$CDM).
The corresponding random errors in the biasing parameters are $5\%$
and $6\%$ for $\Lambda$CDM and $\tau$CDM respectively.

The sampling density can be parameterized by the mean galaxy separation, $l$.  
In our large simulation $l=2.5 \hmpc$, much smaller than the smoothing 
length of $8\hmpc$, but in real samples $l$ could be on the order of the 
smoothing length.  To test the effect of sampling density, we produce 9
mock catalogs in which galaxies are sub-sampled at random from the 
original catalog such that the mean separation is $l=$ 6, 8, or $10 \hmpc$,
while the smoothing length and large volume are kept fixed with
$\neff\sim 3700$. 
The results for $\Lambda$CDM are shown in the bottom panels of \Figu{errors},
and for the two models in Table~2 (columns 5-7).
We see that the sparse sampling artificially enhances both positive
and negative density fluctuations, which enlarges the width of the
galaxy PDF. This results in a steeper biasing function.
For $\Lambda$CDM, the effect becomes noticeable only when $l \geq 8\hmpc$,
where the systematic error in the biasing parameters is of order 10\%
and larger, 
and $\Delta$ is of order a few percent.
For $\tau$CDM the sampling-density effect is noticeable already for $l
\sim 6\hmpc$, 
with the error reaching $30-50\%$ at $\l \sim 10\hmpc$.
A plausible explanation for why the sparse sampling is more damaging in the 
$\tau$CDM model is that the clustering in this model is weaker ($\sigma_8$ is 
smaller to match the cluster abundance which constrains $\sigma_8
\Omega^{0.5}$), 
and therefore the high-density regions are poorly sampled by galaxies.

In summary: the main source of error in our analysis is the sparse sampling. 
For recovering the biasing function with TH8 smoothing, the mean separation 
should be $\leq 8\hmpc$.

\section{CONCLUSION}
\label{sec:conc}
  
We propose a simple prescription for recovering the mean
nonlinear biasing function from a large redshift survey. 
The biasing function is defined by $b(\delm)\,\delm = \av{\delg | \delm}$,
and is characterized to second order by two parameters, 
$\bh$ and $\bt$, measuring the mean biasing and its nonlinearity
respectively. The method is applied at a given cosmology, time, object 
type and smoothing scale, and involves one parameter that should 
be assumed a priori --- the rms mass density fluctuation $\sigma$ 
on the relevant scale.  

The main steps of the algorithm are as follows:
\begin{enumerate}
\item 
Obtain the observed cumulative distribution function in redshift space 
$\cgz(\delgz)$, by counts in cells or with window smoothing at a 
certain smoothing length. 
\item
Assume a value for $\sigm$ on that scale and for 
the cosmological density parameter $\om$,
and approximate the mass CDF in redshift space by $\clnz(\delmz;\sigmz)$, the
cumulative log-normal distribution (\eqd{cln}), with the width $\sigmz$ derived
from $\sigm$ and $\om$ by \equ{sigmz}.
\item 
Derive the mean biasing function by  
\be
\delg (\delm\!=\!\delmz) \simeq
\delgz (\delmz ) = \cgz^{-1} [\clnz(\delmz;\sigmz)] \ \ . 
\ee 
\end{enumerate}

We first showed that the mean biasing function, at TH8 smoothing,
can be derived with reasonable accuracy from the r-space CDFs of 
galaxies (or halos) and mass, despite the biasing scatter. 
We then demonstrated that for a wide range of CDM cosmologies
the mass CDF can be properly approximated for this 
purpose by a log-normal distribution of the same width $\sigm$. 
Next we showed that the biasing functions in z-space and r-space 
are very similar, and that the z-space mass CDF can also be approximated by
a log-normal distribution, with a width derived from $\sigm$ via \equ{sigmz}.
This allows us to apply the method directly to the observed CDF 
in a redshift survey.
The errors in the recovered biasing function and its moments, 
in an ideal case of dense sampling in a large volume, are 
at the level of a few percent.

In any realistic galaxy survey the limited volume and discrete sampling
introduce further random and systematic errors.
For a survey like the PSC$z$ survey, the main source of error is the sampling
density; the error does not exceed $\sim 10\%$ as long as the mean observed
galaxy separation 
is kept smaller than the smoothing radius.
We are currently in the process of applying this method to the PSC$z$
survey (E. Branchini, \etal 2000, in preparation), where a more
specific error 
analysis will be carried out. 
The sampling errors are expected to be significantly smaller
for the upcoming 2dF and SDSS redshift surveys.

In \se{CDF} we showed that our method works well both for halos and
for galaxies, on scales 5 to 15$\hmpc$, and in the redshift range 
$0\leq z \leq 3$ over which the biasing is expected to change drastically.
We obtain a similar accuracy when we vary the cosmological model, 
the mass of the halos in the comparison, or 
galaxy properties such as morphological type and luminosity.
The approximation $\delg(\delta)$ is 
consistent (the deviation is less than 1-$\sigma$) with the true average
biasing function $\coav$  
over a wide range of $\delm$ values, which covers 98 -- 99\%
of the volume, depending on redshift and the type of biased objects.  
This allows us to estimate the moments of the biasing function to
within a few percent (see Table~1).
The moments of the biasing function are derived from 99.9\% of the
volume (99\% at $z$=3 and for relative biasing). 

The method requires as external parameters the rms mass-density
fluctuation $\sigm$ and the cosmological parameter $\om$.
These can be obtained by joint analyses of constraints from
several observational data sets, such as 
the cluster abundance (\eg, Eke \etal 1998), 
peculiar velocities (\eg, Dekel \& Rees 1994; Zaroubi \etal 1997; 
Freudling \etal 1999), 
CMB anisotropies (\eg, de Bernardis \etal 1999),
and type Ia supernovae (Riess \etal 1998; Perlmutter \etal 1999).
Examples for such joint analyses are Bahcall \etal (1999)
and Bridle \etal (1999).

The method is clearly applicable at $z\simeq0$ with available redshift surveys
and especially with those that will become available in the near
future, 2dF and SDSS.
In the future, this method may become applicable at higher redshifts as well,
where the biasing plays an even more important role.
With the accumulation of Lyman-break galaxies at $z\sim 3$, it may soon become
feasible to reconstruct their PDF by counts in cells, and our
method will allow a recovery of the biasing function at this early
epoch, with consequences on galaxy formation and on the evolution of
structure.  

We have concentrated here on smoothing scales relevant to galaxy biasing, 
but the method may also be applicable for the biasing of galaxy clusters,
on scales of a few tens of Mpc.
The biasing scatter may be larger for clusters because of their sparse
sampling, but the larger mean biasing parameter for clusters may help
in regaining the required monotonicity for \equ{cc} to provide a valid
approximation to the mean biasing function.
The mass PDF has been checked to be properly approximated by a 
log-normal distribution at smoothing scales in the range 20 to $40\hmpc$,
using simulations of the standard CDM and Cold+Hot DM models
(Borgani \etal 1995). The errors due to sparse sampling would require
a smoothing scale at the high end of this range.

In a large redshift survey which distinguishes between object types,
one can measure the {\it relative} biasing function between two object 
types by applying \equ{cc12} in redshift space, using the 
observed CDFs for the two types without appealing to the underlying
mass distribution at all.  The upcoming large redshift surveys 
2dF and SDSS, and the DEEP survey at $z\sim 1$, are indeed expected to
provide adequate samples of different galaxy types. 
Compared with the predictions of simulations and semi-analytic modeling 
of galaxy formation
(\eg, Kauffmann \etal 1999; Benson \etal 1998; Baugh \etal 1999; 
Somerville \& Primack 1999), the measured relative biasing function
can provide 
valuable constraints on the formation of galaxies and the evolution
of structure.

While implementing the method outlined above for measuring the mean nonlinear
biasing function using current and future redshift surveys,
the next challenge is to devise a practical method for measuring the 
biasing scatter about the mean.

\acknowledgments{We thank S. Cole, A. Eldar, G. Ganon, T. Kolatt, 
R. Somerville and our collaborators in the GIF team,
J.M. Colberg, A. Diaferio, G. Kauffmann, and S.D.M. White, 
for providing simulations and mock catalogs. 
We thank A. Maller, I. Szapudi and D. Weinberg for stimulating
discussions, and V. Narayanan and M. Strauss for a helpful referee report.
EB thanks the Hebrew University for its hospitality.
This work was supported by the Israel Science Foundation grant 546/98.
}



\def\re{\reference}
\def\jeru{in {\it Formation of Structure in the Universe},
     eds.~A. Dekel \& J.P. Ostriker (Cambridge Univ. Press)\ }

\vfill\eject


\bigskip\bigskip
\cl{
\begin{tabular}{cccccccccccc}
\multicolumn{12}{l}{ {\bf Table 1:} Recovery of the
biasing function from the CDFs} \\
\hline\hline
\multicolumn{12}{c}{$\Lambda $CDM}\\
 & \multicolumn{3}{c}{halos vs. mass} & &
\multicolumn{3}{c}{galaxies vs. mass} & &
\multicolumn{3}{c}{early vs. late type} \\
& $z$=0 & $z$=1 & $z$=3 && $z$=0 & $z$=1 & $z$=3 && $z$=0 & $z$=1 &$z$=3\\
$\bh$     &0.67  &1.21  &2.98  &&0.89  &1.31  &2.38  &&1.11  &1.32&1.28\\
$\bha$    &0.58  &1.25  &2.86  &&0.80  &1.32  &2.25  &&1.20  &1.38&1.49\\
$\bt$     &0.74  &1.24  &3.04  &&0.91  &1.31  &2.40  &&1.13  &1.34&1.30\\
$\bta$    &0.75  &1.31  &3.08  &&0.90  &1.36  &2.38  &&1.35  &1.52&1.64\\
$\Delta$  &0.16  &0.14  &0.11  &&0.08  &0.08  &0.08  &&0.55  &0.38&0.56\\
\hline
\multicolumn{12}{c}{$\tau$CDM}\\
 & \multicolumn{3}{c}{halos vs. mass} & &
\multicolumn{3}{c}{galaxies vs. mass} & &
\multicolumn{3}{c}{early vs. late type} \\
& $z$=0 & $z$=1 & $z$=3 && $z$=0 & $z$=1 & $z$=3 && $z$=0 & $z$=1 &$z$=3\\
$\bh$    &0.90  &2.18  &6.62  &&0.93  &1.71  &4.44  &&1.17  &1.34  &1.27\\
$\bha$   &0.89  &2.28  &6.75  &&0.93  &1.75  &4.32  &&1.18  &1.39  &1.50\\
$\bt$    &0.93  &2.20  &7.85  &&0.95  &1.71  &4.62  &&1.18  &1.35  &1.31\\
$\bta$   &0.96  &2.30  &8.00  &&0.98  &1.76  &4.63  &&1.26  &1.46  &1.65\\
$\Delta$ &0.08  &0.07  &0.20  &&0.08  &0.04  &0.08  &&0.22  &0.20  &0.54\\
\hline\hline
\end{tabular}
}

\bigskip\bigskip\bigskip\bigskip
\cl{
\begin{tabular}{cccccccccc}
\multicolumn{8}{l}{ {\bf Table 2:}
Redshift distortions and sampling errors in the biasing function } \\
\hline\hline
\multicolumn{8}{c}{$\Lambda $CDM}\\
 & True & z-space & z-space ln & Volume & $l=6^a$ & $l=8^a$ & $l=10^a$ \\
$\bh$    &1.13
&$1.12 \pm 0.006$
&$1.09 \pm 0.02$
&$1.12 \pm 0.05$
&$1.17 \pm 0.05$
&$1.23 \pm 0.04$
&$1.31 \pm 0.06$   \\
$\bt$    &1.14
&$1.12 \pm 0.006$
&$1.10 \pm 0.02$
&$1.13 \pm 0.05$
&$1.17 \pm 0.05$
&$1.24 \pm 0.04$
&$1.32 \pm 0.06$ \\
$\Delta$ & ---  
&$0.001 \pm 0.001$
&$0.002 \pm 0.001$
&$0.005 \pm 0.006$
&$0.006 \pm 0.006$
&$0.016 \pm 0.010$
&$0.049 \pm 0.028$  \\
\hline
\multicolumn{8}{c}{$\tau $CDM}\\
 & True & z-space & z-space ln & Volume & $l=6^a$ & $l=8^a$ & $l=10^a$ \\
$\bh$    &1.188
&$1.18 \pm 0.002$
&$1.11 \pm 0.02$
&$1.21 \pm 0.06$
&$1.35 \pm 0.07$
&$1.55 \pm 0.07$
&$1.80 \pm 0.07$ \\
$\bt$    &1.192
&$1.18 \pm 0.002$
&$1.11 \pm 0.02$
&$1.21 \pm 0.06$
&$1.36 \pm 0.07$
&$1.55 \pm 0.07$
&$1.81 \pm 0.07$ \\
$\Delta$ & ---  
&$0.002  \pm 0.0003$
&$0.016  \pm 0.011$
&$0.072  \pm 0.063$
&$0.177  \pm 0.178$
&$0.563  \pm 0.368$
&$1.505  \pm 0.564$ \\
\hline\hline
\multicolumn{8}{l}{$^a$ in units of $\hmpc$ }
\end{tabular}
}

\end{document}